# Absence of dipolar ordering in Co doped CuO

N Vijay Prakash Chaudhary[1], J. Krishna Murthy,[1] A.Venimadhav[1]*

[1]*Cryogenic Engineering Center, Indian Institute of Technology, Kharagpur -721302, India.*

## Abstract

Polycrystalline CuO samples with Co doping were prepared by solid state method with flowing oxygen condition and examined their structural and multiferroic properties. Structural studies have confirmed single phase monoclinic crystal structure of all samples, however, in Co doped samples a decrease in volume with an increase in monoclinic distortion is found. For pristine sample, temperature dependent magnetization has confirmed two antiferromagnetic (AFM) transitions at 213 K and 230 K and frequency independent dielectric peaks at these AFM transitions suggesting the ferroelectric nature. Magnetization of the Co doped samples has showed a marginal increase in ordering temperature of the high-temperature AFM transition and decrease in low temperature AFM ordering temperature. Further, doped samples have shown giant dielectric constant with no signature of ferroelectricity. The x-ray photoelectric spectroscopy study has revealed multiple valance states for both Co and Cu in the doped samples that simultaneously explain the giant dielectric constant and suppression of ferroelectric order.

---------------
***Corresponding author:** venimadhav@hijli.iitkgp.ernet.in*

**Introduction**

CuO is an exceptional member of the rocksalt series of 3d transition metal monoxides [1]. It has a monoclinic structure as opposed to other monoxides with rocksalt cubic structures and exhibited substantially lower Neel temperature ($T_N$) [1, 2]. Since the discovery of high-temperature superconductivity in the copper oxide perovskites, CuO has gained a significant attraction in modern solid state materials research. The Jahn-Teller nature of $Cu^{2+}$ ion is another exception in CuO compared to other rocksalt series of 3d transition metal monoxides. Though understanding of magnetic property of bulk CuO is complex, nevertheless, a renewed interest is seen in exploring transport, magnetic, and optical investigations, especially in view of newly developed nano-materials and various applications [3].

CuO shows two antiferromagnetic (AFM) transitions below room temperature, a low temperature commensurate (CM) transition at $T_{N1} \sim 213$ K and a high temperature incommensurate (ICM) transition at $T_{N2} \sim 230$ K [4]. Below $T_{N1}$, in the AF phase, the magnetic moments are aligned collinearly along b, while above $T_{N1}$ till $T_{N2}$ the magnetic moments are in a spiral state with half of the magnetic moments in the ac plane. Multiferroic behavior has been found in CuO single crystal with ferroelectricity having magnetic origin [2], but, only recently, magnetoelectric effect is reported at large magnetic fields [5]. Effect of doping on the dipolar ordering has been discussed from first principle study and preliminary experiments by Hellsvik et al [6] has shown that 5 % of Zn doping in CuO single crystal retains the ferroelectric ordering, however, the CM and ICM transitions decreases with doping. Recently, hydrostatic pressure on CuO has been predicted to enhance both $T_{N2}$ and the ferroelectric ordering [7].

In single crystal multiferroic CuO a small dielectric permittivity of the order of ~10 is found, while in non-multiferroic CuO polycrystals a giant dielectric permittivity is observed due to the existence of micro traces of $Cu^{3+}$ [8]. In terms of magnetic transitions, Li doping in CuO has resulted in decrease of $T_{N1}$ with substantial increase in conductivity [9, 10]. In search of dilute magnetic semiconductors, ferromagnetism was reported in Mn and Fe doped CuO [11, 12]. Here we report, multiferroic behavior in CuO polycrystalline sample prepared by co-precipitation method and calcined and sintered in oxygen environment. Under similar condition, CuO doped with cobalt (Co) a non-Jahn-Teller cation, showed volume reduction and decrease in low temperature magnetic transition with completely suppressed dipolar ordering.

**Experimental**

Polycrystalline samples of $Cu_{1-x}Co_xO$ (x = 0, 1, 3 and 5 %) were prepared by co-precipitation method. Pure $Cu(NO_3)_2 \cdot 3H_2O$ and $Co(NO_3)_2 \cdot 6H_2O$ were taken in a required molar ratio and dissolved in distilled water; excessive NaOH solution was added to the solution rapidly so that all the metal ions get precipitated. The obtained precipitate was washed with distilled water to remove the redundant $OH^-$. The resultant precipitate was dried at 80 °C to get the precursor powders; then it is calcined at the 1000 °C for a duration of 10 h in oxygen flow and again it is grinded and sintered in the same condition. The phase is identified by High Resolution X-ray diffraction (HRXRD) data collected from Panalytical X'pert PRO PW 3040/60 X-ray diffractometer using the Cu Kα radiation (λ = 1.5418 Å). The valence states of the sample have been determined using X-ray photoelectron spectroscopy (XPS) recorded on a PHI 5000 VersaProbeII system. For dielectric measurements, silver paste was coated uniformly on both sides of the pellet and then measured using precision LCR meter (HIOKI 3532-50). The magnetization measurement of the samples is carried using a SQUID magnetometer (Quantum Design, USA) in the temperature range from 5 - 380 K.

**Result and discussion**

The high-resolution X-ray diffraction (HRXRD) data of all samples are shown in Fig. 1(a) and the structure is analyzed by Rietveld refinement using FullProf Suite. Typical Rietveld refinement of the experimental data (of pristine CuO) is given in Fig. 1(b). The crystal structure of all the samples has been assigned to monoclinic with space group C2/c. The lattice parameters from Rietveld refinement is tabulated in Table 1. With Co doping, the value of lattice parameters changes in all directions, while the overall volume decreases with doping. Considering close matching of the ionic radii of Co with Cu, considerable change in lattice parameters is not expected, however, removal of Jahn-Teller distortion with Co can largely influence the structural parameters and this indeed true and can be found in table 1. But the increase in the angle 'β' indicates the increase in monoclinic distortion. This is surprising, because, in spite of replacing the Jahn Teller distortion with Co, the structure prefers monoclinic with more distortion. In fact, it is documented that the monoclinic structure is very stable even up to high pressures [13]. Since the magnetic ordering involved has superexchange coupling between $Cu^{2+}$ ions; in fact the atomic structure of CuO can be described as connected networks of zigzag Cu-O chains oriented along the [10$\bar{1}$] and [101] directions [14]. From the refinement, we have noticed that the Cu-Cu distance along [10$\bar{1}$] increases while [101] decreases.

The temperature dependent magnetic measurements of the pristine CuO with zero field cooling (ZFC) and field cooled (FC) magnetization in 100 Oe is shown in the Fig. 2(a). Magnetic susceptibility shows a subtle change with two kinks $T_{N1}$ and $T_{N2}$ [2, 15]. The Fig. 2(b) shows temperature dependent real part of dielectric constant, ε' as a function of temperature for the prisitine CuO. We found a low $\varepsilon_r$ value of 9.5 and nearly temperature independent properties down to 100 K. The $\varepsilon'$ behavior shows two kinks at 230 ($T_{N1}$) and 213 K ($T_{N2}$) that corresponds to ICM and CM magnetic transitions and the position of the kinks are frequency independent (see inset of Fig 2 (b)). This observation is similar to the ferroelectric single crystals [2], indicating ferroelectric order in polycrystalline CuO sample.

In case of doped samples, for better comparison magnetization is normalized with magnetization at 300 K and is plotted against temperature in Fig. 3 (a). Two transitions are clearly visible in all the Co doped samples. It can be observed that the high temperature AFM transition can be identified with the $T_{N2}$ of the parent and the position is slightly higher at 232 K. Previously, a small doping with Fe in CuO has also shifted $T_{N2}$ to lower temperature [16]. In case of Li doping, only one magnetic transition was observed and that was too shifted to the lower temperature with doping [9]. In our study, a large difference can be notice with the second transition in the doped ones. The second transition temperature has decreased to 180 K compared to 213 K of pristine sample. However, the sharp rise in magnetization of the second transition in doped samples resembles increased FM interaction than AFM ordering. From the plot, it is clear temperature region between $T_{N1}$ and $T_{N2}$ transitions is widened compared to parent one in a way similar to the observation made by Hellsvik et al [6]. The magnetization is enhanced with the increase of Co concentration. The MH measurements of the samples are shown in the Fig. 3(b), where the hysteresis loop with small coercivity is found and the overall moment increases with Co doping. The increased monoclinic distortion as well as increased magnetization with Co doping is intriguing. However, considering the higher total spin (S) of Co (for either +2 or +3 valances) compared with Cu can be an obvious reason for the magnetization enhancement.

Further, the inhomogeneous superexchange interaction due to different magnetic ions can also give rise to higher magnetization due to competing magnetic interactions.

The dielectric behavior of doped samples (x = 1, 3 and 5 %) were found to be entirely different. The Fig. 4 (a) and (b) shows typical temperature dependent ε' and Tan δ (dielectric loss) of 1 % sample. A clear relaxation behavior is observed above 125 K and it is shifted towards the high-temperature side with increasing of frequency. The Fig. 4 (c) shows ε' of 1 %, 3 % and 5 % doped samples, the relaxation temperature shifts to high temperature with doping, further, the absence of frequency independent peaks indicates the suppression of ferroelectricity. Interestingly, the dielectric permittivity of the doped samples increases by hundred folds. This suggests the possibility of electrically heterogeneous system, this behavior can occur due to the crossover in the relaxation time associated with the grains and grain boundaries. We have analyzed the relaxation mechanism by the thermally activated Arrhenius behavior given by

$$\tau(T) = \tau_0 \exp\left(\frac{E_a}{k_B T}\right) \tag{1}$$

where $E_a$ is the activation energy required for the relaxation process and $\tau_0$ is the pre-exponential factor. The plot of τ versus 1/T is shown in Fig. 4 (d) where the solid lines indicate the fit to equation (1) for x = 1, 3 and 5 % respectively. The obtained activation energy and relaxation time are 0.162, 0.135 and 0.125 eV and $4.62 \times 10^{-16}$, $1.22 \times 10^{-13}$ and $1.55 \times 10^{-12}$ s for 1, 3 and 5 % samples respectively. The relaxation time of 1 % sample is rather too fast and needs to be looked in by other possible conduction mechanisms. It suggests that with the increase in doping, high dielectric loss and grain boundary together can lead to the giant $\varepsilon_r$ and possible Maxwell-Wagner effect. The giant value of $\varepsilon_r$ and high loss for the x = 1, 3 and 5 % samples suggests the presence of mixed valances in the doped samples [8].

In order to understand the absence of ferroelectric nature and large dielectric permittivity of doped samples, we have carried XPS measurement. The core-level XPS of Cu and Co transition metals is measured at room temperature as shown in Fig. 5 (a), (b) and (c). Fig. 5(a) of pristine CuO shows $Cu^{2+}$ $2p_{3/2}$ and $2p_{1/2}$ with binding energies (BEs) of 930.6 and 950.1 eV respectively, indicates the single $Cu^{2+}$ state.[17] The 5 % doped samples shown in Fig. 5 (b) & (c) shows Co $2p_{3/2}$ and Cu $2p_{3/2}$ respectively. The 2p core-level spectrum of Cu $2p_{3/2}$ peak consisting of two peaks at the BEs of 930.6 and 931.6 eV respectively indicates the presence of 2+ and 3+ states in the doped sample. Similarly, the 2p core-level spectrum of Co $2p_{3/2}$ peak consisting of two peaks at the BEs of 780.2 eV and 778.5 eV corresponding to $Co^{3+}$ and $Co^{2+}$ valance states.[18] This study infers that Co doping in CuO induces the mixed valence states in both Co and Cu. This corroborates with the absence of ferroelectric order and high dielectric permittivity of the doped samples. With four valances states all contributing to magnetic exchange interactions, one can expect competing magnetic interactions leading resultant canted magnetic behavior that could explain the higher magnetization in the doped samples.

**Conclusion**

In summary, we have prepared polycrystalline CuO and $Cu_{1-x}Co_xO$ (x = 1, 3 & 5 %) samples by co-precipitation method. The multiferroic nature is confirmed in the pristine polycrystalline sample with frequency independent peaks at magnetic ordering temperatures and very low dielectric permittivity at low temperatures. Co doping preserves the monoclinic structure while decreasing the volume of the cell. Magnetization has improved with Co doping however, the low temperature magnetic ordering has decreased significantly; while the ferroelectric behavior is suppressed in spite of preparing in oxygen environment due to the formation of mixed valence states of both Cu and Co ions.

Acknowledgements

The authors acknowledge DST- FIST facility in Cryogenic Engineering Centre and IIT Kharagpur funded VSM SQUID magnetometer.

Fig. 1 (a) Experimental HRXRD data of x = 0, 1, 3 & 5 % samples and (b) Rietveld refinement of pristine CuO.

Fig. 2 (a) ZFC and FC curves of pure CuO sample with inset shows the region of two AF transition, and (b) Real part of dielectric permittivity with inset showing anomaly at two AF transitions.

Fig. 3. For doped samples (a) shows temperature variation of ZFC-FC magnetization and (b) shows MH curves.

Fig. 4. (a) Temperature variation of $\varepsilon'$ and (b) Tan $\delta$ in CuO for 1 % sample. (c) shows $\varepsilon'$ of a plot of 1, 3 & 5 % samples at 10 kHz, and (d) $\tau$ versus 1/T plot of 1, 3 & 5% samples and the solid line indicates the Arrhenius fit to the experimental data.

Fig. 5. Core level XPS spectra for (a) Cu 2 p for x=0 %: (b) Cu 2 p and (c) Co 2 p core-level for x = 5 %

Table 1: Lattice parameters derived from Rietveld refinement

| Amount of Doping | a (Å) | b (Å) | c (Å) | β (°) | V (Å$^3$) |
|---|---|---|---|---|---|
| **0 %** | 4.6906(4) | 3.4171(3) | 5.1301(5) | 99.374(1) | 81.14 |
| **1 %** | 4.6997(3) | 3.4089(2) | 5.1307(4) | 99.706(1) | 81.02 |
| **3 %** | 4.7141(5) | 3.3907(4) | 5.1275(6) | 99.910(2) | 80.74 |
| **5 %** | 4.7253(1) | 3.3706(1) | 5.1231(2) | 100.193(2) | 80.31 |

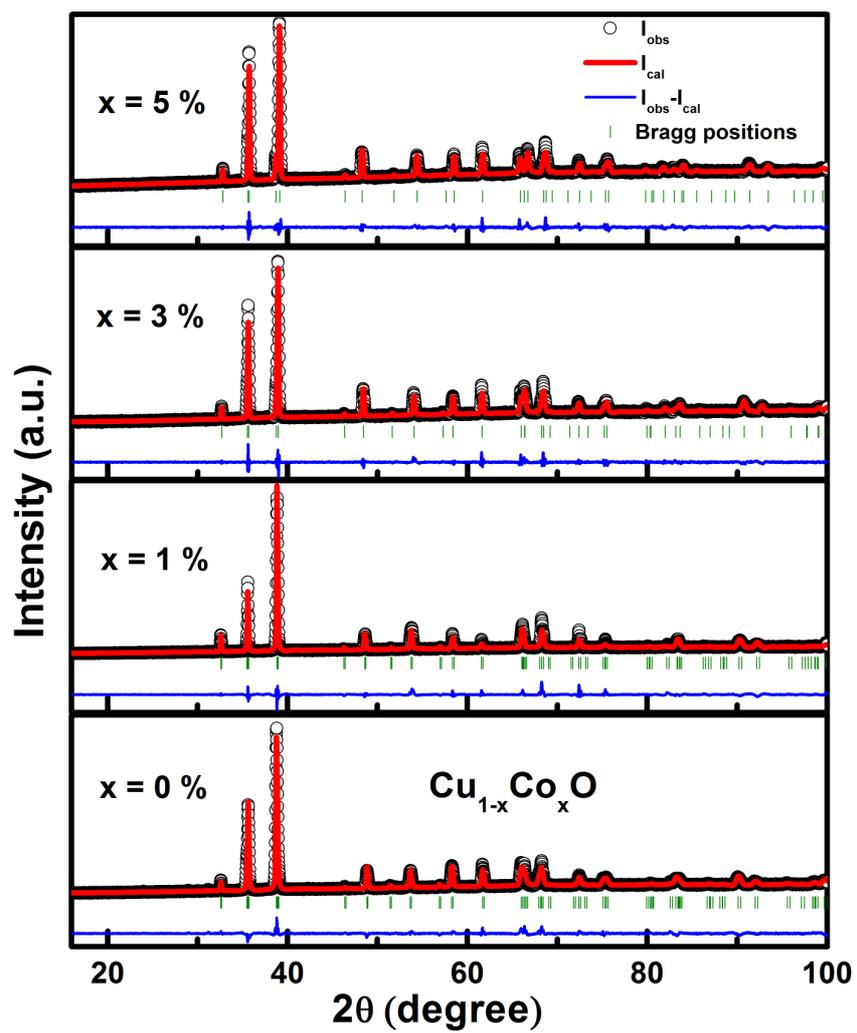

Fig 1

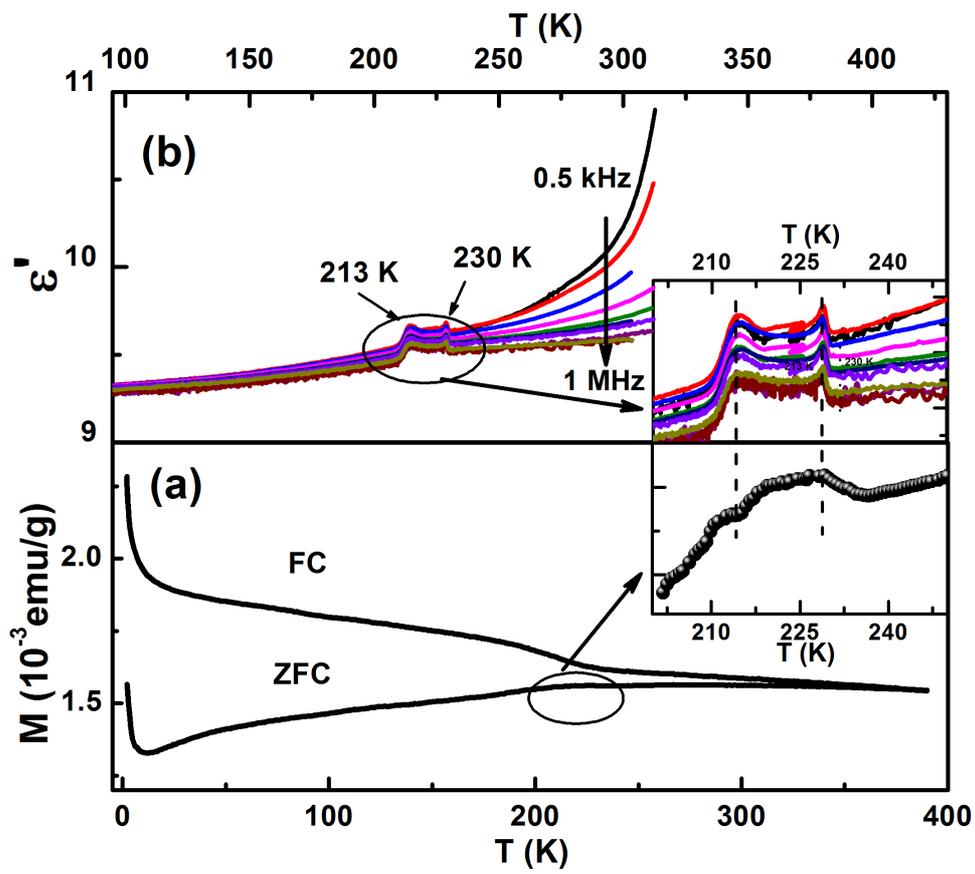

Fig 2

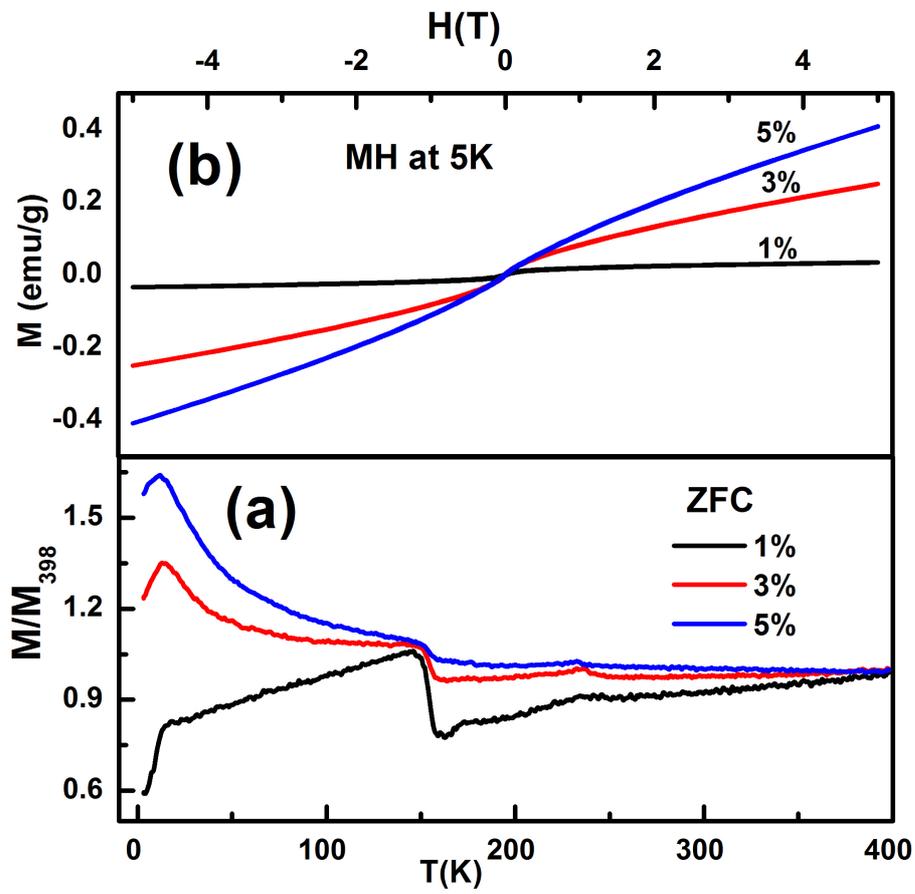

Fig 3

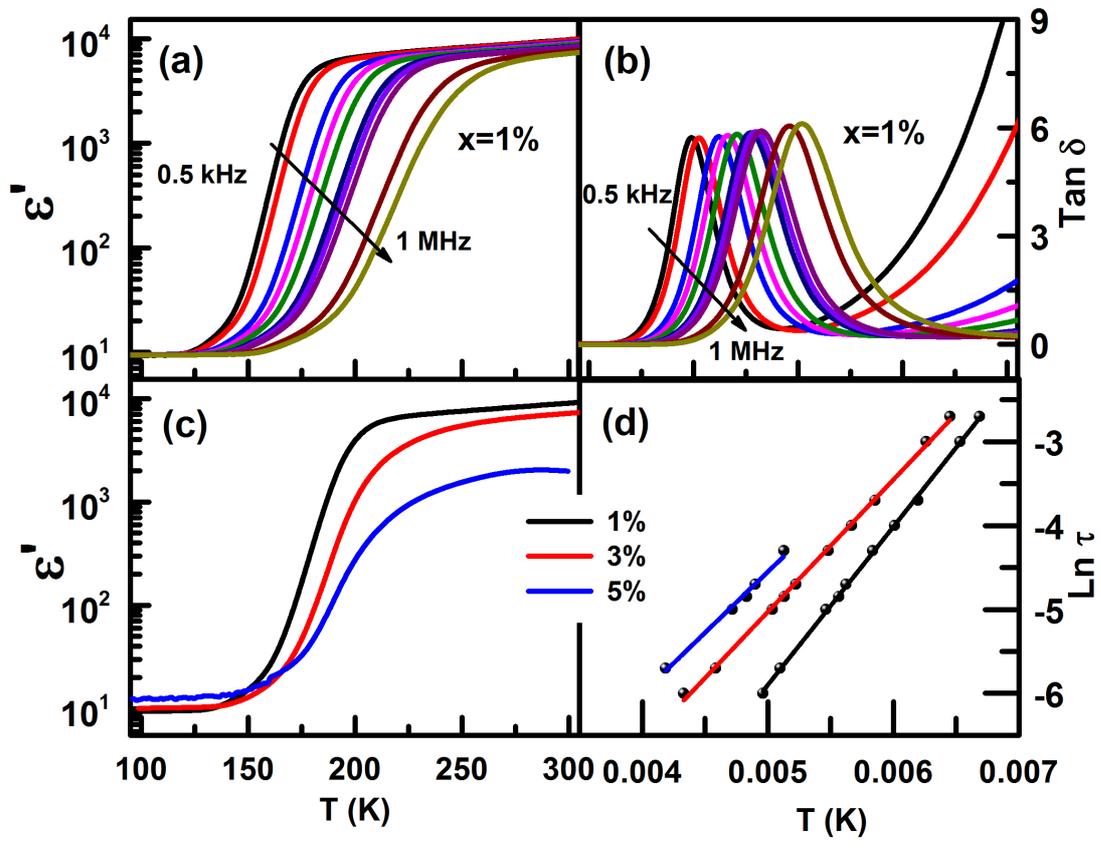

Fig 4

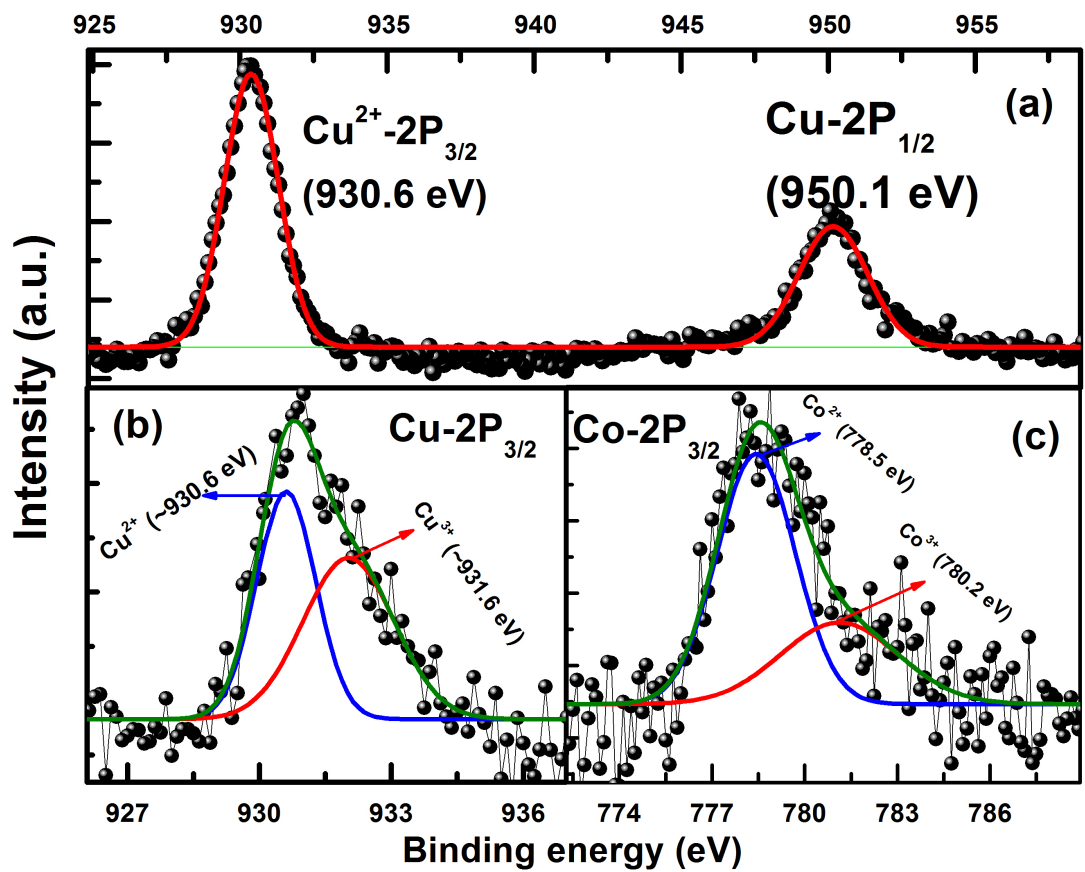

Fig 5